# Single-Ion Sensing in Liquid Using Fluorescent h-BN Point Defects

Yecun Wu[1], Kun Xu[2], Hori Pada Sarker[3], Takashi Taniguchi[4], Kenji Watanabe[4], Frank Abild-Pedersen[3], Arun Majumdar[2,5], Yi Cui[5,6,7], Yan-Kai Tzeng[8], Steven Chu[1,5,9]

1. Department of Physics, Stanford University, Stanford, CA, USA
2. Department of Mechanical Engineering, Stanford University, Stanford, CA, USA
3. SUNCAT Center for Interface Science and Catalysis, SLAC National Accelerator Laboratory, 2575 Sand Hill Road, Menlo Park, CA, USA.
4. Research Center for Electronic and Optical Materials, National Institute for Materials Science, Tsukuba, Japan
5. Department of Energy Science and Engineering, Stanford University, Stanford, CA, USA
6. Department of Materials Science and Engineering, Stanford University, Stanford, CA, USA
7. Stanford Institute for Materials and Energy Sciences, SLAC National Accelerator Laboratory, 2575 Sand Hill Road, Menlo Park, CA, USA.
8. Applied Energy Division, SLAC National Accelerator Laboratory, 2575 Sand Hill Road, Menlo Park, CA, USA.
9. Department of Molecular and Cellular Physiology, Stanford University, Stanford, CA, USA

Corresponding authors:
Yan-Kai Tzeng (ytzeng@slac.stanford.edu), Steven Chu (schu@stanford.edu)

**Abstract**

Understanding the chemical state of individual ions in solutions is crucial for advancing knowledge of complex systems. However, sensing systems at the single-ion level in liquid environments remains a significant challenge. A strategy is introduced that leverages the optical emission properties of point defects in hexagonal boron nitride (h-BN) as single ion sensors. The interaction of optically active h-BN defects with ions in solution leads to distinct spectral shifts, enabling precise visualization and analyzing of individual ions. Using $Li^+$ ions in organic electrolytes as a model, spectral shifts exceeding 10 nm were observed upon ion addition. Application of an external electric field further enhanced these shifts to over 40 nm, enabling real-time monitoring of electrical field induced local perturbations of $Li^+$ ions. Through this approach, individual point defects were shown to spectroscopically distinguish ions of varying charges (e.g., $Na^+$, $Mg^{2+}$, and $Al^{3+}$) based on their local electrical field, each producing a distinct spectral shift. This platform allows direct sensing of ions and their chemical states in liquid environments, providing insights into subtle interfacial changes at the single-ion level, with measurable spectral shifts detectable at millisecond temporal resolution and at concentrations down to 10 micromolar range. This capability presents potential applications in various fields involving ions in liquids that include battery technology and environmental science.

## Introduction

The ability to sensing the chemical state at the smallest scale, *i.e.*, single ion level, is critical across a wide range of scientific and technological fields. Traditional methods for studying the materials at small scales combine microscopy with spectroscopy. Microscopy visualizes the material, while spectroscopy provides chemical information. For example, transmission electron microscopy (TEM) paired with electron energy loss spectroscopy (EELS) can visualize materials at the atomic level and reveal chemical states.[1–3] However, TEM imaging conditions are often incompatible with ionic reaction, such as chemical reactions or electrochemical processes, which require ambient pressure and minimal beam damage. Additionally, the time-consuming procedures and sophisticated instrumentation further limit TEM's potential. Other techniques, like synchrotron X-ray[4] microscopy, atomic force microscopy (AFM)[5] integrated with Raman spectroscopy,[6] or infrared spectroscopy,[7] are constrained by ensemble averaging and cannot resolve individual behaviors. Electronic and electrochemical sensors with ion selective function are the methods of choice for trace bulk quantification and mapping mesoscale reactivity.[8,9] They offer excellent limits of detection (below micromolar) and high-throughput readout, but by design they report spatially averaged activities/potentials or device-scale currents and generally do not reveal which ion occupies which nanoscale site or its residence time at the liquid–solid interface.

Given these limitations, there is a pressing need for innovative approaches that can provide rapid, sensitive, and selective sensing of chemical state at the smallest scale in diverse liquid environments without disturbing the systems of interest. This is where advancements in materials science, particularly the use of h-BN with optically active point defects, present a promising alternative. The single point defect in h-BN has been demonstrated to be optically active as a quantum emitter.[10–13] Recent studies using h-BN point defects in liquid employed optically detected magnetic resonance to probe ensemble responses to paramagnetic spins.[14,15] Besides the quantum information, with single molecule localization microscopy (SMLM),[16–19] the trajectories of protons or organic molecules can be extracted, providing deeper insights into their dynamics.[20,21] In addition, first-principles calculation shows that vacancy sites stabilize $Li^+$ relative to pristine regions.[22] The study of defects in h-BN also has relevance in semiconductor technology, since h-BN is considered as an atomically flat insulating substrate.[23] Nevertheless, the spectral information, which reflects the local electronic structure associated with the defects and indicates the chemical behavior of the surrounding environment, remains unexplored.

Here $Li^+$ interactions with point defects in h-BN in various electrolytes were initially explored. By utilizing SMLM combined with a multispectral localization technique, a red shift in the emission wavelength of over 10 nm was detected upon the addition of $Li^+$ ions to the electrolytes. This spectral shift is attributed to the adsorption and occupancy of $Li^+$ ions together with their solvating electrolytes at defect sites, consistently observed across the three different electrolyte environments studied. The presence of adsorbed $Li^+$ modifies the local electrostatic environment, thus altering the electronic band structure at these defect sites. Applying voltages resulted in a more pronounced red shift of over 40 nm, indicating a de-solvation process. These voltage-induced spectral shifts are also relevant for electrochemical studies involving h-BN.

Extending from this understanding, we demonstrate an effective method to visualize and differentiate solvated ions using single-defect emission spectroscopy from h-BN. Although oxygen plasma treatment creates diverse defect configurations on the h-BN surface, leading to broad features of the ensemble spectral, individual defects consistently exhibit narrow and stable emission spectra. By leveraging this intrinsic stability, we show that solvated ions adsorbed onto single defects can be distinguished clearly based on their ionic charge states. Experimentally, ions with identical charges (e.g., $Li^+$, $Na^+$, $K^+$, $Cs^+$) exhibit similar spectral characteristics. In contrast, ions with differing charges (e.g., $Na^+$, $Mg^{2+}$, $Al^{3+}$) produce distinct spectral shifts that correlate positively with increasing ionic charge. This charge-dependent spectral shift is attributed to stronger local electrostatic fields from higher-charged ions, which significantly perturb the defect's electronic levels, resulting in measurable redshift. These findings uncover fundamental interactions between point defects and solvated ions, advancing single-defect spectroscopy in h-BN as a next-generation tool for label-free, charge-sensitive ion detection at the nanoscale.

This innovative approach holds the potential to transform a range of fields involving ions in liquid. In energy storage research, its ability to spectroscopically resolve the behavior of individual ions provides unprecedented insight into key battery processes, from electrolyte interaction to ion dynamics, including de-solvation, intercalation, dendrite growth, and solid-electrolyte interphase (SEI) formation.[24] For example, batteries fail or age from rare, highly localized interfacial events; tens-of-nanometer, single-ion mapping lets us find those hot spots (where ions first accumulate, de-solvate, or react), which bulk electrochemical measurements average out. By enabling a deeper understanding of these phenomena, it offers a pathway toward the development of next-generation batteries with improved cyclability and higher capacity. The applicability demonstrated in both organic and aqueous electrolytes further open opportunities to explore a broad spectrum of battery systems, such as alkali metal and aqueous batteries. Combining high sensitivity, charge selectivity, and real-time monitoring, this technique emerges as a powerful, label-free tool for probing reaction processes. Ultimately, it enables direct sensing of ions with their chemical state within a liquid environment with millisecond temporal resolution and single-ion sensitivity.

## Main

Point defects in h-BN can be intentionally induced into pristine material through large-area irradiation with ions or neutrons, or by oxygen plasma etching, as outlined in the Methods section. With plasma etching, these point defects become discernible on the surface of h-BN, as shown through atomic resolution scanning transmission electron microscopy (STEM) in Fig. 1a, using integrated differential phase contrast (iDPC) mode. The point defects in h-BN are optically active, as schematically shown in Fig. 1b. When lithium ions in electrolyte approach and bind to these defects, they induce measurable changes in the emission properties (Fig. 1c). Monitoring these changes enables the effective visualization of individual lithium ions at the point defects within the electrolyte.

The plasma-treated h-BN was fabricated into a microfluidic cell for imaging as shown in Extended Data Fig. 1 (see Methods). At the initiation of the experiment, there is an excess of emitters, which may stem from impurities and unstable defects on the surface of h-BN. Upon sample illumination, the emitted light underwent a photobleaching process before reaching a steady state as shown in Extended Data Fig. 2. Following photobleaching, the number of emitters per frame remained stable for observation times beyond 150 seconds, using 20 ms exposure at a 10 Hz frame rate.

Figures 1 d depict the consecutive super-resolution images of emitters on the same flake in the steady state over 1500 frames in pure propylene carbonate (PC) electrolyte and PC electrolyte with 1 M $Li^+$ ion. The corresponding video is shown in Supplementary Video 1. The statical data of the number of emitters (*N*) during the steady state in three different lithium-ion electrolyte (PC, fluoroethylene carbonate FEC, dimethyl sulfoxide DMSO) are shown in Extended Data Fig. 2, as well as the photon counts. It is evident that the number of emitters increased by more than fivefold in all three electrolytes. This increase occurs alongside the presence of lithium ions in the electrolytes, and the distribution of the number of emitters conforms to the Poisson distribution. Both prior experiments and simulations support that cations preferentially interact with defects in h-BN.[20,22]

A concentration-dependent experiment was performed on a different flake, as shown in Fig. 1e. At concentrations below $10^{-6}$ M, the results become indistinguishable due to background emission from pre-existing emitters, impurities, or minor contamination introduced during the experiment, thereby defining the detection limit. At concentrations above $10^{-3}$ M, *N* approaches saturation near unity, which arises from the optical resolution limit and the average ion spacing on the h-BN surface.

The adsorption of solvated ions on defects can be described by the Hill–Langmuir model,[25] neglecting ion–ion interactions since the average ion separation exceeds the Debye length. Because multi-ion binding requires the simultaneous presence of several ions, the defect occupancy rate scales with higher powers of concentration, with the exponent corresponding to the number of ions involved. The relationship between surface coverage and concentration can thus be expressed as:[25,26]

$$\theta(c) = \frac{c^n}{K_d + c^n}, \log\frac{\theta(c)}{1-\theta(c)} = \log K + n \log c \quad (1)$$

where $K_d$ is the dissociation constant, and the Hill coefficient *n* reflects the number of ions bound per defect. Fitting the curve in the intermediate concentration range ($10^{-5}$–$10^{-3}$ M) yields a Hill coefficient of ~0.82, consistent with single-ion binding behavior.[27] The coefficient being less than unity is likely due to partial loss of emitters that became permanently inactive after prolonged measurements.

To investigate the underlying interaction between $Li^+$ ion and point defects in h-BN, a multispectral localization approach was employed to simultaneously image the position and spectral characteristics of the emitters (see Extended Data Fig. 3 and Methods).[28] Figure 2a presents emission spectral data from single emitters measured at identical locations on the same flakes, comparing conditions with and without lithium ions in the electrolyte. Notably, the presence of lithium ions in the electrolytes induces a redshift of over 10 nm in emission wavelength. This observation is further confirmed in Fig. 2 d-e by statistical analysis of single-defect emissions collected across entire flakes in electrolytes both without and with $LiClO_4$, suggesting the influence of the ionic environment on defect states. This shift could be attributed to the physical adsorption between solvated $Li^+$ ion and point defects, which will be discussed in detail in a later section. The increase in emission intensity after ion incorporation can be understood as a combination of charge-state stabilization, suppression of nonradiative pathways, and possible local solvation/electrical field effects. Mobile ions can electrostatically stabilize bright charge states, adsorbates may passivate nearby traps or compensate charges, and local field effects can reduce spectral diffusion. While the precise microscopic origin remains an open question, similar activation has been reported for protonated and solvent-immersed h-BN emitters.

The device was then modified to incorporate electrodes (see Methods) to confirm that the observed emission shift results specifically from solvated $Li^+$ ions rather than from anions or organic solvent molecules. This modification allows for the application of an electric field to control the ion movement and distribution. Figure 3 a and b schematically illustrate the process: applying a positive voltage to the copper input electrode ($V_{input}$) relative to the indium tin oxide (ITO) creates an electrolyte double layer,[29] leading to the accumulation of lithium cations ($Li^+$) on the surface of the h-BN (Fig. 3a). Conversely, a negative voltage at the copper electrode repels lithium ions from the h-BN surface (Fig. 3b).

The $Li^+$ ion interactions were modulated by applying a square wave voltage to the input electrode ($V_{input}$) with a cycle period of 30 seconds. The number of emitters per frame, shown as yellow dots in Fig. 3c, rises and falls in sync with the square wave transitions. At -1 V, when anions accumulate on the surface of the h-BN, the emitter's counts per frame are low. Conversely, at +1V, lithium ions accumulate on the surface, causing a significant increase in the emitter number. This indicates that the emission is primarily activated by $Li^+$ ions, rather than other components in the electrolyte. The output waveform of the number of emitters resembles a square wave output from a first-order RC low-pass filter. By fitting the observed curve with exponential functions defined by the equations

$$N = P\left(1 - e^{-((t-t_0)/\tau)}\right) + Q \tag{2}$$

for the rising phase and

$$N = Pe^{-((t-t_0)/\tau)} + Q \tag{3}$$

for the falling phase, the steady-state number of emitters at -1 V ($Q = N(t \to \infty) \approx 15$) and +1 V ($P + Q \approx 63$) can be extracted as shown in Fig. 3d, along with the time constant $\tau \approx 6\ s$. This analysis yields a quantitative measure of the emitter dynamics under varying voltage conditions, details of which will be elaborated upon in subsequent sections.

Figure 3e shows super-resolution reconstructed images of emitters captured during the final ten seconds of each of three square-wave pulses. By identifying defect positions relative to the flake edges, emission from the same defect sites can be consistently monitored under varying electrolyte voltages. Emission spectra from a representative defect site (highlighted by white dashed circles in Fig. 3e) during the final ten seconds of each voltage pulse is shown in Fig. 3f. At –1 V bias, some emitters remain visible with spectra consistent to those observed under floating conditions (no applied bias, Fig. 2f). In this regime, solvated cations can still physically be absorbed at defect sites, although the number of active emitters decreases due to electrostatic depletion of cations in the double layer. This apparent similarity arises because, even under negative bias, thermal fluctuations and crowding effects allow a small population of cations to persist near the interface. In contrast, a significant redshift is observed at +1 V, shifting the emission peak to approximately 650 nm.

To further investigate dynamic behavior, a sinusoidal voltage waveform (±1 V amplitude) was applied across two electrodes, with frequencies ranging from 0.02 Hz to 5 Hz (Extended Data Fig. 4a). At lower frequencies, the number of emitters oscillated in sync with the voltage. As the frequency increased, the amplitude of these fluctuations diminished, becoming negligible above 1 Hz. Fourier analysis of both the input voltage ($V_{input}$) and the emitter response revealed that the electrolyte double layer functions as a low-pass filter (Extended Data Fig. 4b), amplifying low-frequency signals while attenuating those above a ~1 Hz cutoff, where noise becomes dominant.

Considering possible reactions between the ITO substrate and $Li^+$ ions, passivated cells were fabricated by coating the ITO with a photoresist layer, which had an opened window to selectively expose only the h-BN flake to the electrolyte, as illustrated in Extended Data Fig. 5 and b. Although boron vacancies can exist in negative charge states,[30] full electron transfer to neutralize $Li^+$ is energetically unfavorable without a conduction pathway and would be expected to yield large changes in the electronic structure. Instead, DFT study show that $Li^+$ remains near the +1 state when adsorbed,[31] with only partial charge redistribution between the defect and the ion. The slow scan exhibits a current obscured by noise, while the fast scan demonstrates capacitive behavior without any detectable Faradaic current (Extended Data Fig. 5c), indicating no electrochemical reactions were observed on the surface on the h-BN.

Ideally, if $Li^+$ ions are to be inserted into defect sites within h-BN from a bottom contact, electrons must also have access to those sites. However, the wide band gap of h-BN (~5.9 eV) results in extremely poor electronic conductivity, making this process highly unlikely under standard conditions. Electrochemical reactions require the simultaneous transport of electrons and $Li^+$ ions to active sites, which depends on a continuous electronic conduction pathway. Since bulk h-BN lacks such pathways, even dense surface defects typically cannot support meaningful charge transfer from a bottom contact. While defect states may locally narrow the band gap or enable hopping or tunneling conduction, these mechanisms are generally insufficient to sustain appreciable current unless the h-BN layer is extremely thin, heavily defective, or deliberately engineered to improve conductivity.

As a result, electrochemical lithium insertion into surface defects through a bulk insulating h-BN layer is not feasible in our system. Likewise, other surface redox reactions such as solid-electrolyte interphase (SEI) formation are unlikely. In contrast, partial or even full de-solvation of $Li^+$ ions can occur at the surface of h-BN without requiring significant redox activity. In solution, $Li^+$ ions are typically surrounded by a solvation shell of solvent molecules. However, as they approach a surface, especially under an applied electric field, they can shed part of this shell to interact with available surface sites. On an insulating substrate like bulk h-BN, which lacks electronic conductivity, $Li^+$ ions cannot be reduced to metallic lithium ($Li^0$) and therefore cannot form stable intercalation compounds. Instead, they may physically or weakly chemically absorb onto defect sites, depending on the local surface chemistry and electrostatic environment. This interaction may be accompanied by partial de-solvation, driven by perturbations in the solvation equilibrium at the interface, even though true electrochemical insertion remains improbable in the absence of a conductive pathway. Specifically, voltage displaces nearby anions and modifies the local electric field at the defect but does not reduce cation–cation separation or promote ion clustering, which would require several-eV driving energies. Thus, the redshifts are attributed to perturbations from single de-solvated ions rather than multi-ion binding.

These findings indicate that even without significant faradaic currents, the system remains sensitive to interfacial processes like $Li^+$ de-solvation and weak adsorption at defect sites. This suggests that the h-BN interface can serve as a platform for investigating non-faradaic or perturbation-driven interfacial reactions, in which variations in local solvation, ion distribution, or surface polarization give rise to measurable optical signatures. As a result, subtle interfacial reactions–often hidden in standard electrochemical measurements–can be observed and characterized. Early studies have demonstrated modulation of single-photon emission via electric fields;[32,33] however, the emitters were embedded within the h-BN rather than located on its surface, and the observed effects primarily arose from carrier injection or tunneling, rather than direct interactions between ions and defects.

Density Functional Theory (DFT) with hybrid function calculations were used to explore the electronic band structure of h-BN. As shown in the calculated density of states (DOS) in Extended Data Fig. 6a, pure h-BN has a wide bandgap. The valence band is mainly made up of nitrogen *p* orbitals, while the conduction band mostly comes from boron *p* orbitals. The study focused on common defects created by oxygen plasma treatment,[34] such as vacancies where boron ($V_B$) or nitrogen ($V_N$) atoms are missing, or where both are missing together ($V_{BN}$). Oxygen plasma treatment primarily introduces O-related defects, which remain polar and chemically active coordination sites for cations.[35,36] Here we included oxygen-related defects like oxygen substituting boron ($O_B$) or nitrogen ($O_N$), and combinations like $V_BO_N$ (boron vacancy with oxygen-nitrogen substitution) and $V_NO_B$ (nitrogen vacancy with oxygen-boron substitution).

Calculations showed that several of these defects ($V_B$, $V_{BN}$, $O_B$) create electronic states within the bandgap, some of which can transition at around 2 eV–potentially leading to fluorescence, as illustrated in Extended Data Fig. 6 b-h. However, the transition in $O_B$ is less likely due to spin mismatch, which costs extra energy for spin flipping. These mid-gap states are key sources of light emission. We also note that the DOS calculation is not able to differentiate between a direct and indirect bandgap. Thus, the identification of our experimentally determined spectra to these DOS interpretation is ambiguous. Additional potential defect configurations involving three or more atoms, such as $V_{BBN}$, $V_{BNN}$, $O_BV_{NN}$ and others, were not calculated. In fact, h-BN samples treated with oxygen plasma likely have a wide variety of local defect environments including carbon related defects. These can also be influenced by things like strain in the material, changes in charge states of defects, or the presence of nearby impurities. All these factors contribute to the broadening of the photoluminescence spectra when measured across many defects at once. The atomic identity of visible h-BN emitters remains under active investigation, with credible assignments spanning boron vacancies, oxygen- and carbon-related substitutions and complexes. Our spectra ~600 nm are consistent with prior reports of carbon-containing centers as well,[37] while oxygen plasma treatment could also produce O-containing or mixed C/O complexes at vacancies. Hydrocarbon contamination and beam-induced carbon deposition in 2D materials complicate unambiguous detection of carbon, and we therefore avoid over-interpretation. Future correlative studies (such as scan tunneling microscopy) may further refine structural attribution.

Notably, restricting the analysis to single, isolated defect sites largely eliminates variability, yielding sharper and more reproducible emission spectra. Adsorption of external ions to these sites, through either physical or chemical interaction, induces modifications to the local electronic environment. These perturbations alter the defect energy levels and produce measurable shifts in the emitted light. Accordingly, h-BN provides a viable platform for ion detection through spectroscopic monitoring of fluorescence shifts.

For $Li^+$, $Na^+$, $K^+$, and $Cs^+$, the emission peak positions shift only slightly toward longer wavelengths as ion mass increases, as shown Extended Data Fig. 7a. Despite nearly a 20-fold increase in mass from $Li^+$ to $Cs^+$, the total shift is only about 6 nm. This suggests that emission wavelength is not strongly affected by ion mass. In contrast, as shown in Fig. 4 a-c, ions in the same periodic row show more distinct spectral features. For example, $Na^+$ emission peaks are centered around ~575 nm, $Mg^{2+}$ near ~585 nm, and $Al^{3+}$ close to ~600 nm. These clear differences highlight the potential to distinguish ions based on their charge states. The full width at half maximum (FWHM) distributions are presented in Extended Data Fig. 7b and Extended Data Fig. 8a, with an average FWHM of around 20 nm.

To reduce variability caused by different defect types, the same ion solution at a single defect site was further analyzed emission spectra. Spectra collected from the same defect location at different times were grouped for comparison. In Fig. 4d–f, the solid line indicates the excellent match, where peak centers align exactly with the group mean values. The dashed lines represent ±3 nm deviations, corresponding to the resolution limit of our spectrometer. Emission peaks falling within this range are considered indistinguishable. These results show that individual emission spectra from the same ion interacting with the same defect remain stable over time, demonstrating consistent and reproducible optical responses. The accumulation of individual spectrum within a group improves the signal-to-noise ratio (SNR), thus providing a clearer demonstration of the spectra as shown in Extended Data Fig. 8b.

Therefore, the spectral 'wandering' observed in ensemble spectra originates from structural heterogeneity of oxygen-plasma-induced defects rather than low SNR. At the single-defect level, the emission peaks remain narrow (10–30 nm FWHM), stable, and reproducible, enabling reliable detection of ion-induced shifts exceeding the linewidth.

An h-BN flake was immersed in a solution containing equal concentrations (1/3 M each) of $Na^+$, $Mg^{2+}$, and $Al^{3+}$ ions to differentiate among these ions. These individual emission spectra were collected and plotted the distributions of their peak centers, as shown in Fig. 4h. Three distinct peak groups clearly appeared, each corresponding to one of the ions. However, some emissions signals appeared between these main groups, making it harder to tell them apart. This is likely due to the variety of defect types present in the material.

Even though the tails of the distributions overlap–because different defects can affect the emission–it is still possible to distinguish ions by looking at emissions from individual defect sites. At the single-defect level, emission spectra are much more consistent and stable, unlike the broader and more varied signals seen in ensemble measurements.

To validate this concept, Figure 4g presents emission spectra from the same single defect, which was exposed sequentially to $Na^+$, $Mg^{2+}$, and $Al^{3+}$. As the ion charge increases, the emission shifts gradually from around 575 nm to 600 nm. This clear trend shows that ions with different charges can be reliably identified using single-defect measurements.

The variation of emission signals with ion concentration and exposure time was studied, using $Li^+$ or $K^+$ as representative examples. Extended Data Fig. 9 shows the number of active emitters per frame ($N$) across concentrations from $10^{-3}$ M to $10^{0}$ M under identical acquisition settings. $N$ increases from $10^{-3}$ M to $10^{-2}$ M, followed by a slight downturn toward $10^{0}$ M. This behavior reflects the onset of saturation at $10^{-2}$ M, with the small decrease above this concentration falling within experimental uncertainty. Meanwhile, the per-emitter brightness remains approximately constant across all concentrations. Additionally, the emission peak positions stayed consistent, showing that the spectral response is stable across a wide concentration range. While many electrochemical sensors surpass micromolar bulk limits of detection, our method targets a different metric: single-ion, single-site spectroscopy and dynamics at the liquid–solid interface, including ion differentiation and trajectories, which bulk measurements, by design, do not provide.

Next, the effect of exposure time was explored using a fixed $K^+$ concentration of $10^{-3}$ M, with exposure times ranging from 1 s down to 1 ms. The optimal SNR was obtained at an exposure time of about 10 ms (Extended Data Fig. 10), indicating that $K^+$ ions typically remain at defect sites for only a few milliseconds. Shorter exposures produced weak signals, making the peaks harder to detect. We observed photon counts exceeding $10^4$ photons per second with a 1 ms exposure time and reaching up to $10^6$ photons per second with a 1 s exposure time, which are comparable to, or in some cases surpass, previously reported results.[13,20,38] These results highlight the high sensitivity and fast response of this technique for detecting ions at the single-defect level. Importantly, the observed photon rates are primarily limited by the residence time of the ions on defects, rather than by optical inefficiencies. Although RC charging of the electrochemical double layer exhibits response times on the order of several seconds (Fig. 3c–d, Extended Data Fig. 4), ion differentiation in this system is instead dictated by the millisecond-scale residence times of individual ions at surface defects.

The ion-induced redshift arises from the local ionic environment created by the charged ions, rather than from bulk solvent effects. Further theoretical investigations would be valuable to clarify the mechanism, particularly in disentangling the roles of solvation structure and electrostatic interactions. In terms of electrostatics, a single charged ion located at the length scale of its first solvation shell (~0.3 nm) above the h-BN surface is expected to generate an electric field on the order of ~5 MV $cm^{-1}$, after accounting for both Debye screening and the reduced effective dielectric constant of interfacial water.[39] The measured shifts (10–20 nm, corresponding to tens of meV) fall within the range previously reported for h-BN single-photon emitters under external electrical field,[40,41] supporting the assignment to electrostatic field effects rather than irreversible structural modifications. Beyond that, solvation structure change could also contribute to it. Compared to the spectral changes induced by solvent-related dielectric effects, the bulk electrolyte environment here remains unchanged.[21] However modifications in the local dielectric constant[42] and the ion–sensor separation due to solvation structures are non-negligible and should be taken into account.

Additionally, ion detection does not require strong binding: even transient $Li^+$ residence above a defect produces local electric fields sufficient to induce measurable spectral shifts. Although the precise atomic structures of h-BN emitters remain under debate, the reproducible optical response of single defects to local ionic fields establishes a robust basis for single-ion sensing.

By contrast, ion size and mass mainly affect atomic vibrations and exert only a minor influence on the electronic structure, often through indirect mechanisms such as electron-phonon coupling. This explains why the spectral differences among $Li^+$, $Na^+$, $K^+$, and $Cs^+$ remain relatively small compared with those arising from variations in ion charge (Extended Data Fig. 7). Additionally, a quantitative first principles understanding of the spectral properties of the emission properties of these defect states is beyond the scope of this paper.

The present study focused primarily on alkali and alkaline earth metal ions, while transition metal ions with partially filled 3*d* orbitals, such as $Fe^{2+}$, $Fe^{3+}$, $Cu^{2+}$, and $Mn^{2+}$, were not systematically investigated. These ions often display distinctive colors due to *d*-*d* electronic transitions and charge transfer interactions with the surrounding ligands like water molecules.[43,44] This coloration may interfere with the emission caused by interactions with h-BN defects. For instance, $Fe^{3+}$, which has an outer electron configuration of $3s^2$, $3p^6$ and $3d^5$, showed no observable fluorescence within the current optical detection window. Future studies will focus on exploring the fluorescence behavior of transition metal ions, particularly $Fe^{3+}$, in interaction with h-BN, to better understand how orbital-level effects influence defect emission.

Lastly, three core performance indicators of the platform were evaluated. First, the detection limit extends down to $10^{-5}$ M, constrained by background emitters and impurities, and saturates at concentrations above $10^{-2}$ M. Second, under ambient conditions, temperature primarily modulates kinetic rates but does not measurably affect the ion-dependent spectral shift. The temperature induced shift of h-BN emitters near room temperature is typically negligible (~0.01–0.1 meV/K)[45,46] compared with ion-induced shifts of 10–20 meV. Third, emitter counts plateau after photobleaching for >150 s, defect spectra remain reproducible within ±3 nm, and repeated ±1 V bias cycles yield consistent RC-like responses without evidence of Faradaic currents. Collectively, these characteristics demonstrate concentration-invariant fingerprints, predictable kinetic scaling, and robust stability under operando conditions.

## Conclusion

This study demonstrates a significant advance in ion sensing by using spectral signatures to track single ions in liquids. Monitoring shifts in light emissions as ions interact with their surroundings provides insights into ions behavior and dynamics, especially in systems involving charge carriers and point defects in h-BN.

$Li^{+}$ ions were shown to activate point defects in h-BN, producing redshifts exceeding 10 nm. Application of an external electric field further enhanced these shifts to over 40 nm, enabling precise detection of the local perturbations induced by individual ions. The approach also distinguishes ions of different charges, $Na^{+}$, $Mg^{2+}$, and $Al^{3+}$, through distinct spectral responses, detectable at millisecond timescales and concentrations as low as 10 μM. In contrast to bulk measurements, which are complicated by heterogeneous defect ensembles, this technique leverages sharp and reproducible emission from individual defect sites to resolve ion-specific interactions.

Overall, this spectroscopic platform provides a robust route for probing complex ionic systems and offers broad potential for advancing studies in chemical energy storage and other ion-related processes in liquid environments.

## Methods

Sample preparation:

Clean #1.5 glass coverslips served as the initial substrates, subjected to a sequential washing protocol involving acetone, isopropyl alcohol (IPA), a potassium hydroxide solution, and finally, deionized (DI) water. After cleaning, h-BN flakes were mechanically exfoliated from high quality crystals[47] onto the substrate of coverslips. Exfoliated flakes were initially identified by color contrast under white-light microscopy to select those with a thickness around 20 nm. For flakes characterized by AFM, the measured thicknesses are reported in the figure captions. Flakes that were too thick exhibited strong background emission from embedded intrinsic defects, while flakes that were too thin were often difficult to identify on the coverslip under white light. Surface activation was achieved through oxygen plasma etching performed with a March Instruments PX-250 Plasma Asher, employing a power setting of 75 W and an oxygen gas flow rate of 10 standard cubic centimeters per minute (sccm). Etching durations were 60 seconds for general sample preparation and increased to 120 seconds specifically for samples destined for electrical response analysis. The defects created by oxygen plasma are primarily located on the surface due to the plasma's low energy, leaving them exposed to solvents.

Samples allocated for electrical measurements were further processed by sputtering approximately 180 nm of indium tin oxide (ITO) onto the cleaned coverslips, followed by an annealing phase at 370 °C for one hour. After the mechanical exfoliation and plasma treatment, the construction of the counter electrode was carried out by affixing a polyimide (PI) Kapton film, upon which copper was sputtered, to the opposite side of the ITO-coated coverslip. This setup ensures the application of voltage for electrical measurements while preventing potential short circuits between the copper and ITO layers. The ITO substrate passivation in Extended Data Fig. 5 was performed by UV photolithography of photoresist followed by thermal cross-linking and hardening over 150 °C. It is worth noting that, even after thermal cross-linking baking, the photoresist can still gradually degrade in organic solvents, making it unsuitable as a permanent passivation method.

To construct the polydimethylsiloxane (PDMS) microfluidic channel, a mixture of polydimethylsiloxane (liquid) and a cross-linking agent is poured into a mold created from Kapton tape (~60 μm thick) and then heated to form an elastomeric replica. After the mold is cut and peeled away, a biopsy punch is used to create inlets and outlets for the electrolyte. The PDMS surface is subsequently treated with oxygen plasma and bonded to a glass coverslip. The electrolyte is introduced into the channel within an argon-filled glovebox and sealed with an additional PDMS layer on top. Lithium perchlorate ($LiClO_4$) was used as the lithium-ion source. Chemicals used for electrolytes are purchased from Sigma-Aldrich or Fisher Scientific, with salts purities exceeding 99.99%. A concentration of 1 M $LiClO_4$ was used for imaging in Fig. 1, while a concentration of 0.1 M $LiClO_4$ was used for electrical response analysis in Fig. 3 and Extended Data Fig. 4. For imaging multiple ions, sulfate salts were used, including $Li_2SO_4$, $Na_2SO_4$, $K_2SO_4$, $ZnSO_4$, and $Al_2(SO_4)_3$, each at a cation concentration of 1M in Milli-Q deionized water. The imaging of $H^+$ was using the Milli-Q deionized water directly.[20]

For the TEM imaging, the sample was initially exfoliated onto a PDMS substrate and then transferred to a microporous $SiN_x$ TEM grid using a deterministic transfer method. Prior to the transfer, the $SiN_x$ grid was sputtered with a ~5 nm thick Pd/Au alloy to facilitate charge dissipation during imaging.

Optical setup:
The microscopy setup is schematically shown in Extended Data Fig. 3a. A 532 nm laser was utilized (Cobolt 06-DPL 532nm 200mW laser) for wide-field excitation via an oil immersion objective (Nikon CFI SR HP Apo TIRF 100XC Oil, NA: 1.49). Wide-field excitation was directed to the backplane of the sample with a power density of 4000 W/cm$^2$. After passing through a 200 mm infinity-corrected tube lens (Thorlabs, TTL200MP), the emission signal was collected through the same objective. The signal underwent further refinement through a series of optical filters—including a dichroic beam splitter (Semrock, FF535-SDi01-25x36), a notch filter (Thorlabs, NF533-17), a long-pass filter (Thorlabs, FELH0550), and a short-pass filter (Thorlabs, FESH0800). It was then concentrated using a 300 mm achromatic doublets lens (Thorlabs, AC508-300-AB). After initial focusing, the luminescence was split by a nonpolarizing 50/50 beam splitter (Thorlabs, BSW10R) into positional and spectral channels. Each channel was focused by a 180 mm doublets lens (Thorlabs, AC508-180-AB) and recombined using a knife-edge mirror (Thorlabs, MRAK25-P01) onto the left and right halves of an EMCCD camera (iXon Ultra 897 EMCCD). To disperse the spectral signal, an equilateral prism (PS863, Thorlabs) was positioned at the Fourier plane of the final focusing lens. A series of narrow band-pass filters (Thorlabs, FBH570-10, FBH600-10, FBH650-10, FBH694-10, FBH750-10) and 35 nm fluorescent nanodiamond[48] was used to calibrate the spectra before data collection. During imaging, the EM gain was set to 200, and the frame rate was adjusted to 10 Hz for samples destined for electrical response analysis. For imaging multiple ions in deionized water, the microscopy was set to total internal reflection fluorescence (TIRF) mode to reduce background signal, using a lower excitation power of 1000 W/cm$^2$. Specifically, for better resolution, the spectral data in Fig. 4 were further calibrated using an additional band-pass filter at 635 nm (Thorlabs, FBH635-10).

All measurements were performed at room temperature. Ion adsorption at defect sites is an equilibrium process with only weak temperature dependence over the liquid-phase range. The mechanism of ion-induced spectral shifts based on electrical field is not affected. Thus, the single-ion sensing reported here is stable under ambient conditions, while systematic temperature-dependent studies may further refine the understanding of ion thermodynamics in future work.

Hanbury Brown–Twiss $g^{(2)}$ measurement is not performed in this work because the spectrally resolved wide field SMLM setup is optimized to track many single defects in liquids under time varying voltages, conditions that are incompatible with a confocal HBT geometry given the ms scale ion residence time. The single photon nature of the h BN defects employed here—including plasma generated and liquid activated emitters—has been established in numerous studies reporting $g^{(2)}(0)<0.5$ at room temperature,[21] including in liquid environments. We therefore focus on single defect spectroscopy and provide multiple internal controls (per defect co localization, reproducible peak positions within spectrometer resolution, narrow FWHM) that are only consistent with single defect signals.

Optical data processing:
Two background images were used to subtract the baseline from the raw data. The first, called the absolute background, is an averaged image captured in an area without h-BN samples but under the same excitation conditions. Additionally, a reference background specific to each emitter was used for extracting spectral information. This reference background consists of the average image of ten neighboring frames that do not contain the emitter of interest. The number of emitters and super-resolution reconstruction were determined using the ImageJ plugin, ThunderSTORM.[16] A wavelet filter is applied to each frame, followed by peak fitting using 2D integrated Gaussian functions. The localization uncertainty varies from 4 to 40 nm, averaging around 22 nm. This depends on the standard deviation of the Gaussian fit of emitter's intensity ($\sigma_{PSF}$ = ~288 nm) and the photon counts of each emitter.[49] For spectrum extraction, emitters located at the edges, those overlapping (which required to be at least 50 pixels apart to prevent spectral overlap), and those with insufficiently low intensities were excluded from the analysis. For the spectra data in Fig. 4, peaks were fitted using a single Lorentzian peak from the range between 550 nm to 700 nm.

TEM:
The STEM images were acquired on Thermo Fisher Spectra 300 monochromated, double-corrected scanning transmission electron microscope. The high voltage is 300 kV, and the convergence semi-angle is 30 mrad with sub-angstrom probe. IDPC-STEM images[50] are acquired under ~ 10 pA beam current by using four-segmented DF-S detectors. The camera length is 185 mm. IDPC images are processed using Velox software. The acquired image was digitally processed through a bandpass filter to highlight the lattice spots.

Theoretical computation:
Spin-polarized Density Functional Theory (DFT) with hybrid functionals (HSE06) calculations were conducted using the Vienna ab initio Simulation Package (VASP-5.4.4)[51,52] within the python-based Atomic Simulation Environment (ASE).[53] The Perdew-Burke-Ernzerhof (PBE)[54] functional within the generalized gradient approximation (GGA) described the exchange and correlation interactions. The projector-augmented plane wave (PAW) method with a kinetic energy cut-off of 600 eV was used to treat ion-electron interactions, ensuring well-converged results. The k-point mesh was sampled using the Monkhorst-Pack scheme. To correct for van der Waals interactions, the DFT-D3 method with Becke-Johnson damping was employed.

A (5×5) four-layer h-BN surface containing 200 atoms was utilized, derived from an optimized bulk structure. For bulk calculations, a primitive unit cell and a 7×7×3 k-point grid were used for geometry optimization. In surface calculations, a 20 Å vacuum was applied between repeating slabs to prevent spurious image-charge interactions. The bottom two layers were fixed to the bulk geometry, while the top two layers and adsorbates (Li-ion) were allowed to relax during surface geometry optimization, below a force threshold of 0.01 eV/Å. A 3×3×1 k-point grid was used for surface geometry optimization, and a higher k-point grid for density of states (DOS) calculations. The effects of solvent and solvation structure of the ions are not included in the computation.

Single-to-noise (SNR) ratio calculation:
For each emitter in Extended Data Fig. 10, emitters were first identified by applying a binary threshold to the image. Areas of 16 pixels or larger were classified as a potential emitter. The total photon count for each emitter was obtained by summing all pixel values within the defined emission area. The background photon count for each emitter was estimated by averaging ten neighboring frames that did not contain emitters in the corresponding area. The SNR for each emitter was calculated by subtracting the total background from the emitter intensity and dividing this difference by the standard deviation of the background values within the emitter area. In other words, each point in Extended Data Fig. 10a is calculated by subtracting the data in Extended Data Fig. 10c from Extended Data Fig. 10b and then dividing the resulting difference by the corresponding values in Extended Data Fig. 10d.

**Supplementary Information**
Supplementary Video 1
Continuous video recordings of a flake immersed in pure PC solvent and in PC containing 1 M $LiClO_4$.

**Data availability**
All data needed to evaluate the conclusions are present in the paper and supplementary information. Source data are available from the corresponding author upon reasonable request.

**Acknowledgement**
We thank Prof. Harold Y. Hwang for the discussion. The optical experiments are supported by the Laboratory Directed Research and Development program at SLAC National Accelerator Laboratory, under contract DE-AC02-76SF00515. The device fabrication was supported by the U.S. Department of Energy (DOE), Office of Basic Energy Sciences, Division of Materials Sciences and Engineering under contract DE-AC02-76SF00515. Yecun Wu and Yi Cui acknowledge the partial support from U.S. Department of Energy (DOE), Office of Basic Energy Sciences, Division of Materials Sciences and Engineering under contract DE-AC02-76SF00515. Yecun Wu acknowledges the support from Stanford Energy Postdoctoral Fellowship and the Precourt Institute for Energy. Yan-Kai Tzeng acknowledges the support from U.S. Department of Energy of the Battery500 Consortium program. Hori Pada Sarker and Frank Abild-Pedersen acknowledge the financial support provided by the U.S. Department of Energy, Office of Science, Office of Basic Energy Sciences, Chemical Sciences, Geosciences, and Biosciences Division, Catalysis Science Program to the SUNCAT Center for Interface Science and Catalysis. Kenji Watanabe and Takashi Taniguchi acknowledge support from the JSPS KAKENHI (Grant Numbers 21H05233 and 23H02052) and World Premier International Research Center Initiative (WPI), MEXT, Japan. Part of this work was performed at nano@stanford RRID:SCR_026695.

**Author contributions**

Y.W., Y.-K.T., and S.C. designed the research; Y.W. performed device fabrication, optical measurement, and data analysis; K.X. performed TEM measurement under supervision of A.M. H.P.S. performed theoretical computation under supervision of F.A.P.; T.T. and K.W. supplied the h-BN crystals; Y.C. discussed and assisted in analyzing the data; Y.W., Y.-K.T., S.C. wrote the manuscript with comments from all the authors.

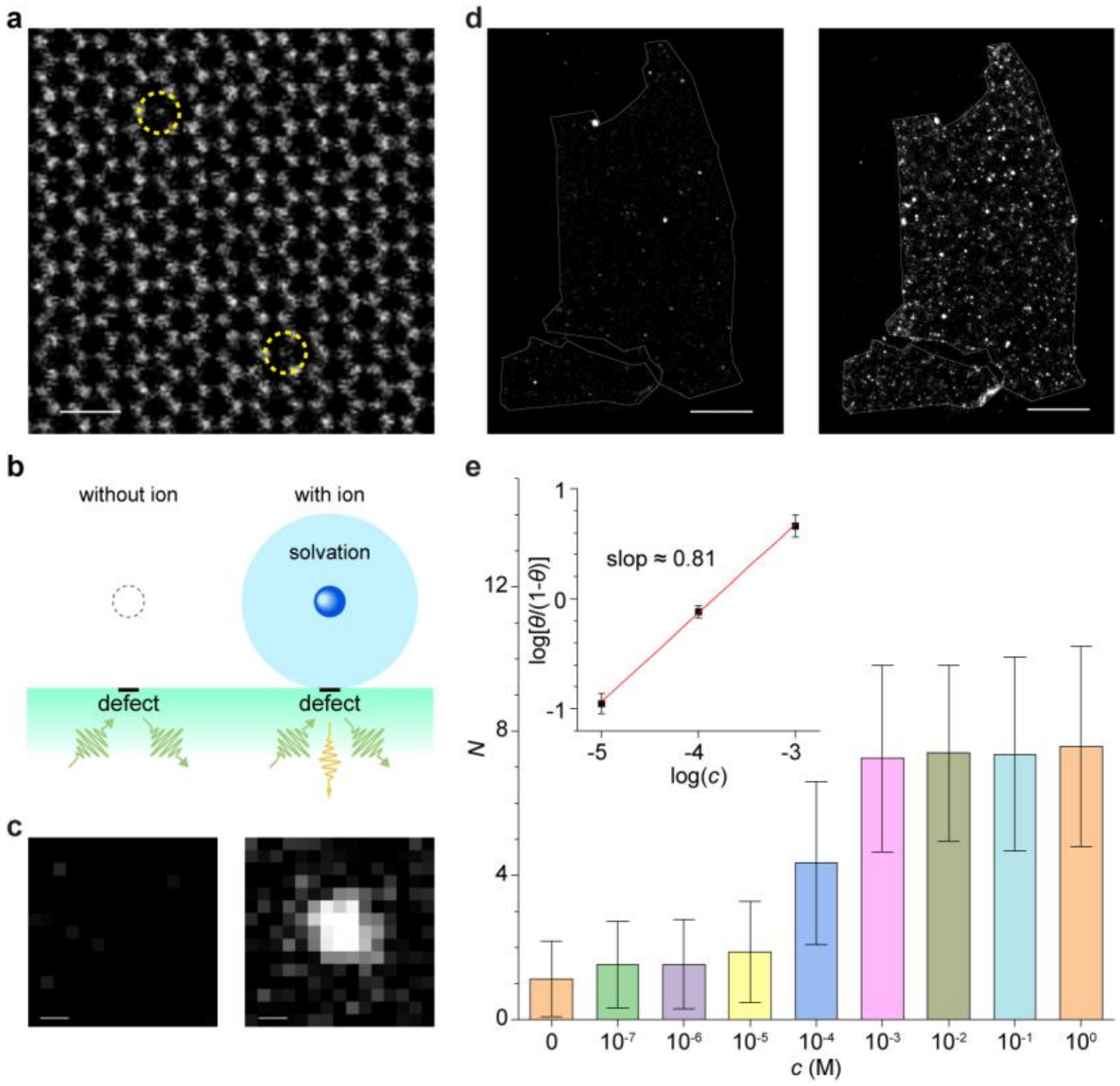

**Fig. 1: Visualization of $Li^+$ activation of quantum defects in h-BN.**

**a**, Integrated differential phase contrast (IDPC)-scan transmission electron microscopy (STEM) image of an oxygen plasma treated few-layer h-BN (scale bar: 500 pm). The yellow dash circles indicate point defect in h-BN. **b**, Schematic representation of the configuration in which an ion occupying the defect site activates emission. **c**, Fluorescence images of the same range on an identical flake without and with $Li^+$ (scale bar: 200 nm). **d**, Reconstructed super-resolution averaged shifted histogram visulization of h-BN flakes in pure PC (left), and 1 M $Li^+$ in PC (right), reconstructed from a sequence of 1500 frames (100 frames per second) in steady state (from 1001 to 2500 frame, data shown in Extended Data Fig. 2). Scale bar: 3 μm. The flake thickness is ~21 nm. **e**, Number of emitters per frame for the same flake in PC solvent at varying $Li^+$ concentrations. For each concentration, data were collected from 100 frames during the steady state. The error bars represent the standard deviation. Inset: logarithmic plot of θ/(1–θ) versus concentration ($10^{-5}$–$10^{-3}$ M), where θ is the surface coverage defined as $[N(c) - N(0)]/N_{max}$, with $N_{max}$ value taken as the average value over $10^{-2}$ – $10^{0}$ M. The error bars represent the standard error by assuming the $N_{max}$ and $N(0)$ are constant. The flake thickness is ~17 nm.

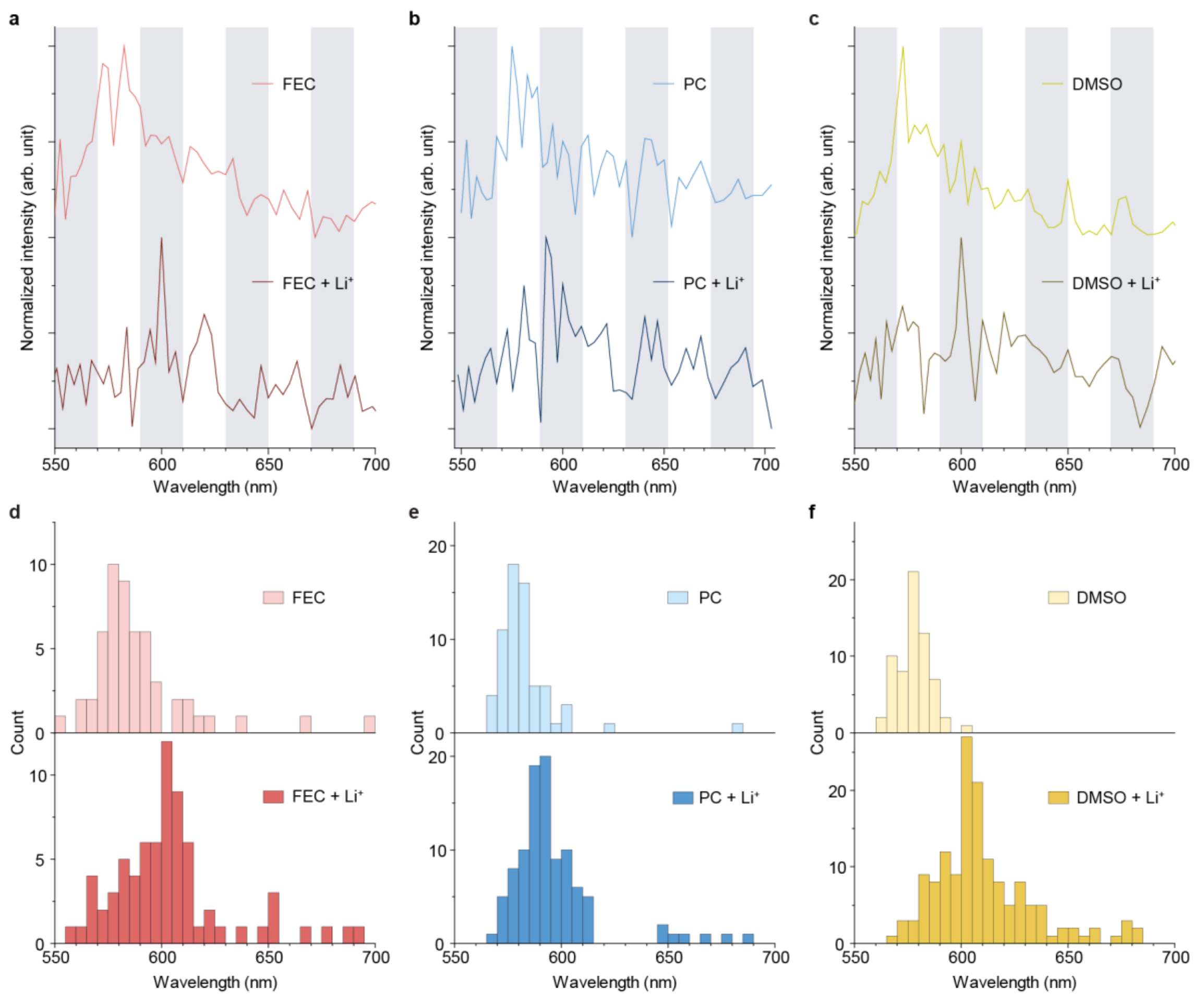

**Fig. 2: Emitter spectra within different electrolytic environments with and without $Li^+$.** **a-c**, Emission spectra from single quantum emitters measured in FEC (**a**), PC (**b**), and DMSO (**c**) electrolytes without and with $Li^+$. Spectra within each panel were collected from the same defects at identical locations. The scatter plots are the raw data, and the solid lines are fitted spectra. **d-f**, Distributions of spectral peak centers for emitters in electrolytes without and with $Li^+$. Within each panel (d, e, f), the distributions with and without $Li^+$ are obtained from the same flake, whereas panels **d–f** represent three different flakes. The bins start at 550 nm with a width of 3 nm, corresponding to the resolution of our spectral extraction.

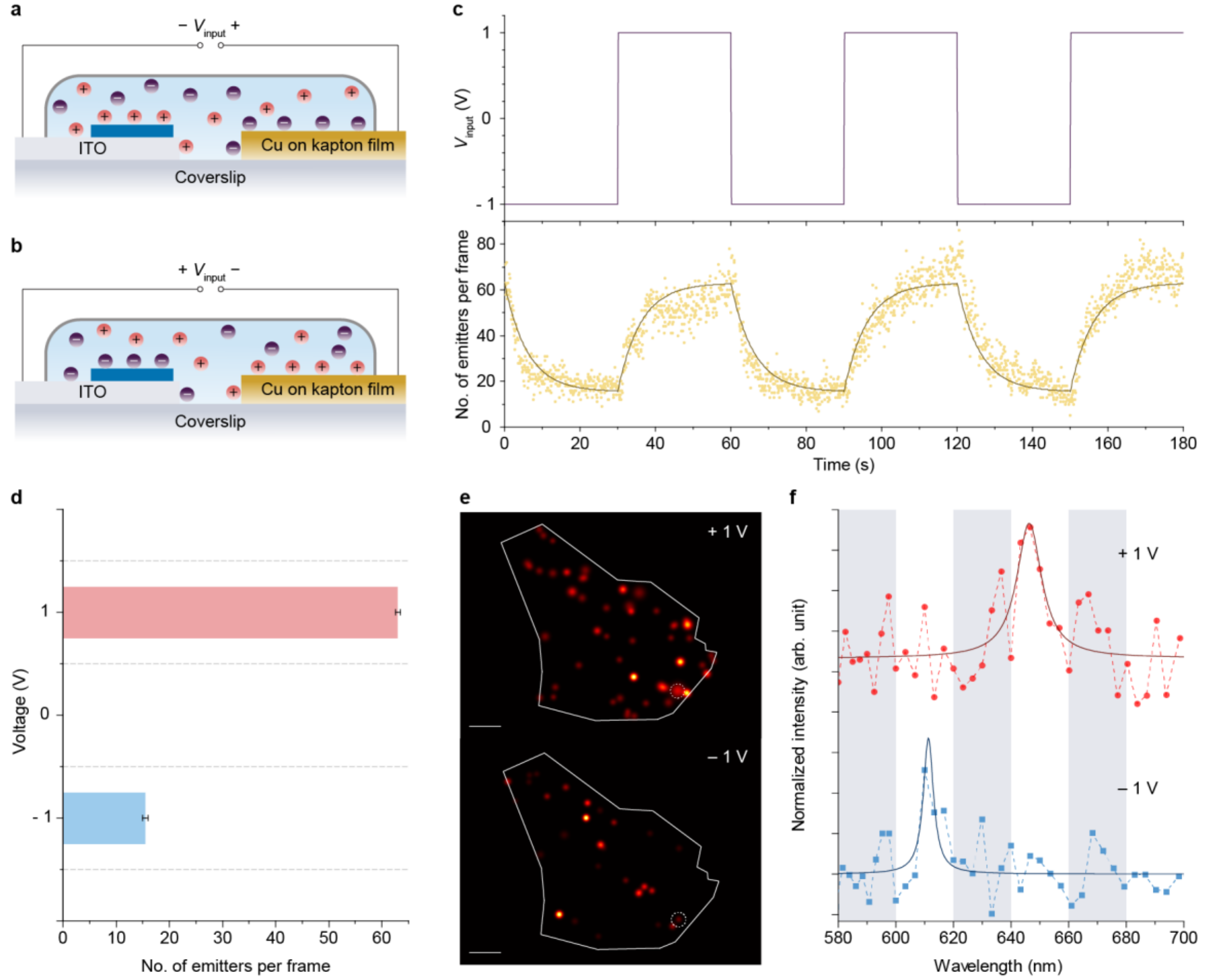

**Fig. 3: Spectra indicated electrochemical reactions on the surface of h-BN.**

**a-b**, Schematic illustration of the cell setup and ion distribution under varying voltages. With a positive $V_{input}$ (**a**), $Li^+$ ions, shown in red, are drawn towards the h-BN surface (shown in blue), creating an electrolyte double layer. Conversely, when a negative $V_{input}$ (**b**) is applied, anions accumulate on the h-BN surface. **c**, Voltage profile of the $V_{input}$ square wave (top, blue) contrasted with the resultant number of emitters fluctuations per frame (bottom, red). The black line in the bottom graph represents the output fitting to a standard RC low-pass filter, modeling the system's response. **d,** The fitted number of emitters per frame at + 1 V and – 1 V respectively. The error bars represent the 95% confidence range of fitting results. **e**, Reconstructed super-resolution images of the h-BN flake by fitting each detected emitter as a Gaussian spot captured at + 1 V (**d**) and – 1 V respectively. The data were collected in the last 10 seconds prior to the voltage switch, specifically at intervals of 20-30s, 80-90s, and 140-150s for -1 V, and 50-60s, 110-120s, and 170-180s for + 1 V. Scale bar: 2 µm. The flake thickness is ~23 nm. **f**, Normalized spectra of an identical spot (dash circles in panel **e**) at + 1 V and – 1 V respectively. The scatter points represent the raw data, while the solid lines indicate the fitted spectra.

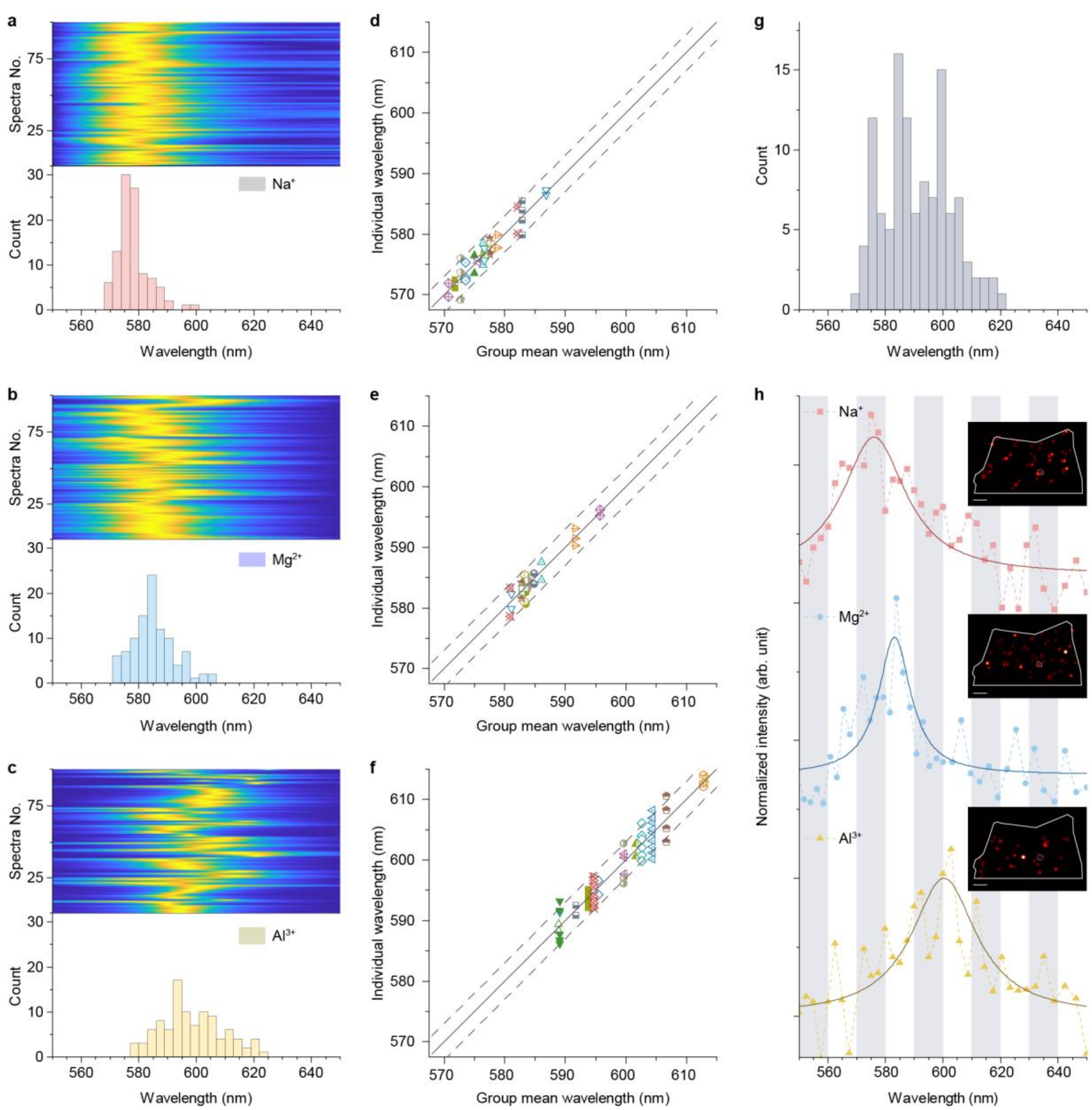

**Fig. 4: Differentiation of multiple ions without external voltage.**

**a–c**, Fitted emission spectra (upper) and corresponding peak-center distributions (lower) from flakes measured in aqueous solutions containing $Na^{+}$ (**a**), $Mg^{2+}$ (**b**), and $Al^{3+}$ (**c**) ions, respectively. Each upper plot shows the top 100 fitted spectra ranked by fitting quality (R-square) and color-coded intensity from blue to yellow. The bins start at 550 nm and have a width of 3 nm, matching the resolution of our spectral extraction. **d–f**, Scatter plots of individual spectral peak-center values versus the group mean wavelength for emitters repeatedly measured at identical positions in $Na^{+}$ (**d**), $Mg^{2+}$ (**e**), and $Al^{3+}$ (**f**) solutions. Different symbols denote different groups of emitters. The solid diagonal line indicates perfect agreement (individual = mean), and dashed lines show a ±3 nm deviation. **g**, Distribution of spectral peak centers from emitters measured on a flake immersed in a solution containing an equal concentration (1/3 M) of $Na^{+}$, $Mg^{2+}$, and $Al^{3+}$ ions. **h**, Representative spectra of the same defect measured separately in $Na^{+}$, $Mg^{2+}$, and $Al^{3+}$ solutions. Solid curves represent the fitted profiles. The flake thickness is ~27 nm. Insets show reconstructed super-resolution images obtained by fitting each emitter as a Gaussian spot; white dashed circles mark the defect positions used for spectral extraction (scale bars: 2 μm).

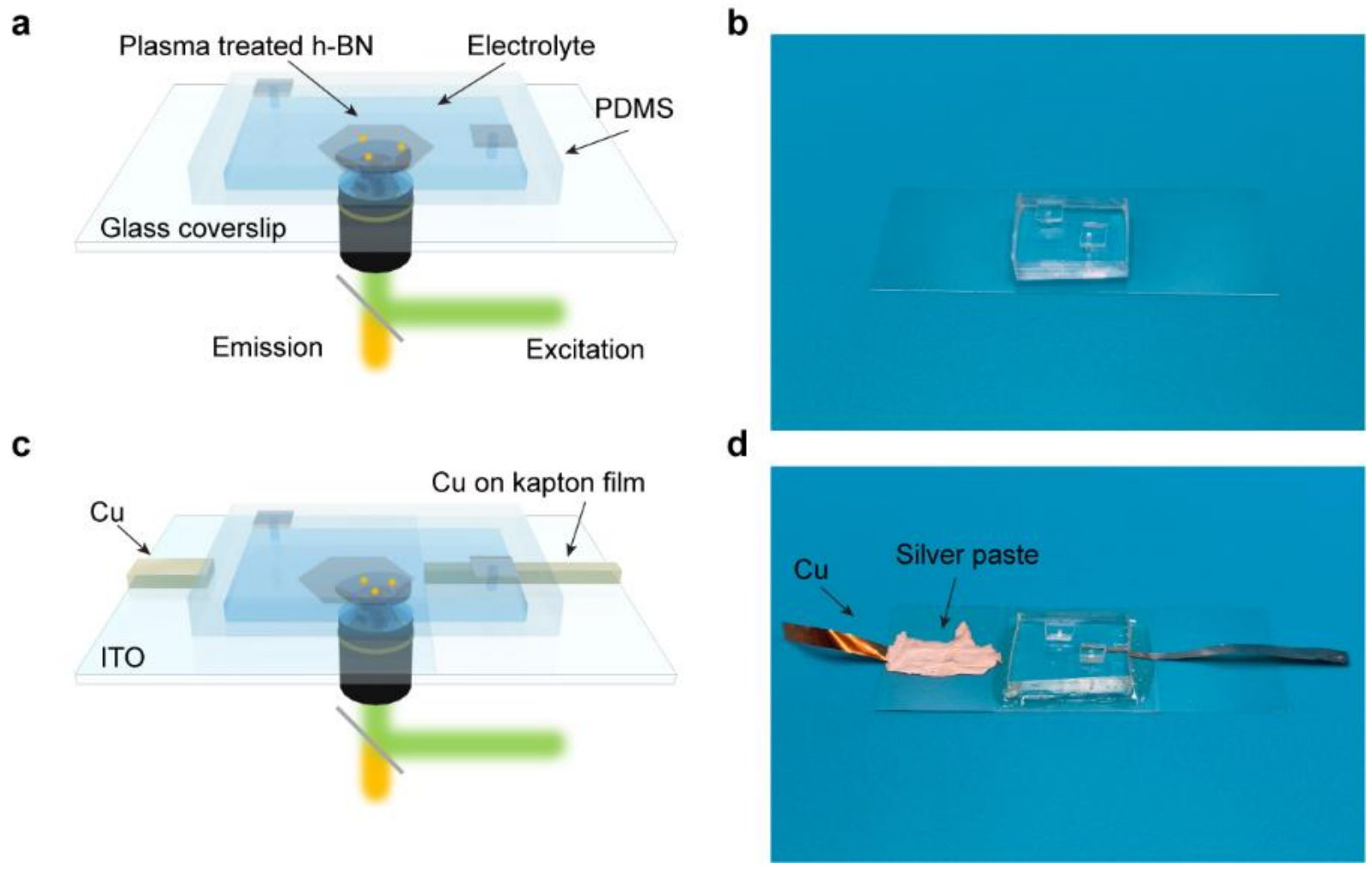

**Extended Data Fig. 1: The schematic and photographs of the microfluidic cell.**
**a-b**, Schematic illustration (**a**) and photograph (**b**) of a standard imaging microfluidic device. **c-d**, Schematic illustration (**c**) and photograph (**d**) of a microfluidic device equipped with electrodes for electrical testing.

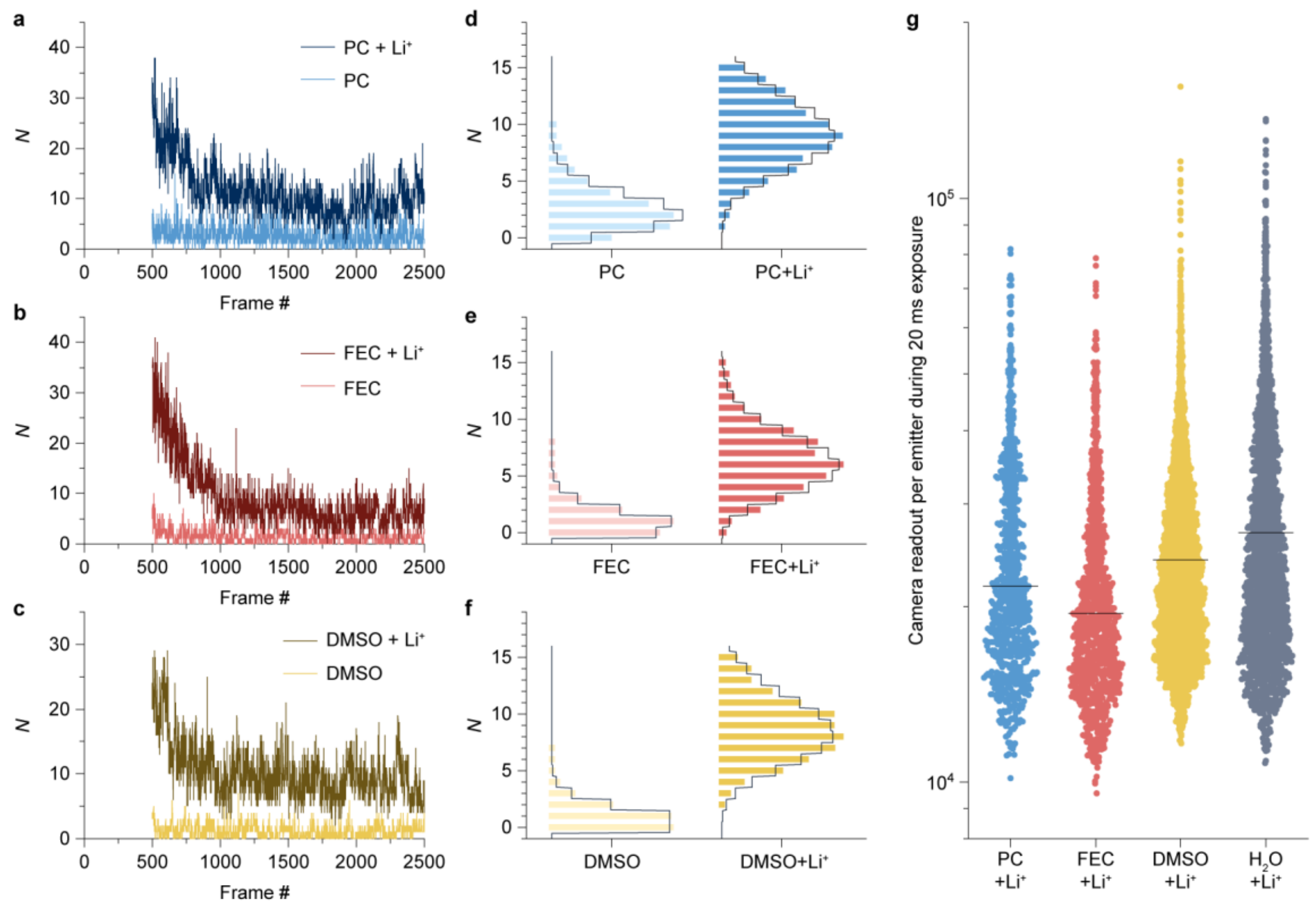

**Extended Data Fig. 2: The number of emitters as a function of illumination time and photon count per emitter.**

**a-c,** The number of emitters per flake stabilize at a steady state after 1000 frames (100 ms per frame) in different electrolytes, both with and without lithium. Before reaching 500 frames, the high density of emitters combined with the manual focus process makes it impossible to accurately count the number. The flake thicknesses from panel **a** to **c** are: 21 nm, 30 nm, and 31 nm respectively. **d-e,** Statistical distribution of the number of emitters per frame from the sequence of 1500 frames in panels **a-c**, in different electrolytes, both without and with 1M $LiClO_4$. The curves are fitted with Poisson distribution. **g,** Readout counts from camera per emitter recorded with a 20 ms exposure time for $Li^+$ ions in different solvents under epifluorescence imaging mode with excitation power density of 4000 W/cm$^2$ and EM gain of 200 during data collection. The $Li^+$ ion source is 1M $LiClO_4$ for FEC, PC, and DMSO, and 0.5 M $Li_2SO_4$ for DI water. The black lines are the median value. Only bright emitters, representing the top 1% of pixel counts in the entire image, are included in the analysis. Dim emitters, which are not suitable for spectral extraction, are excluded from the plot.

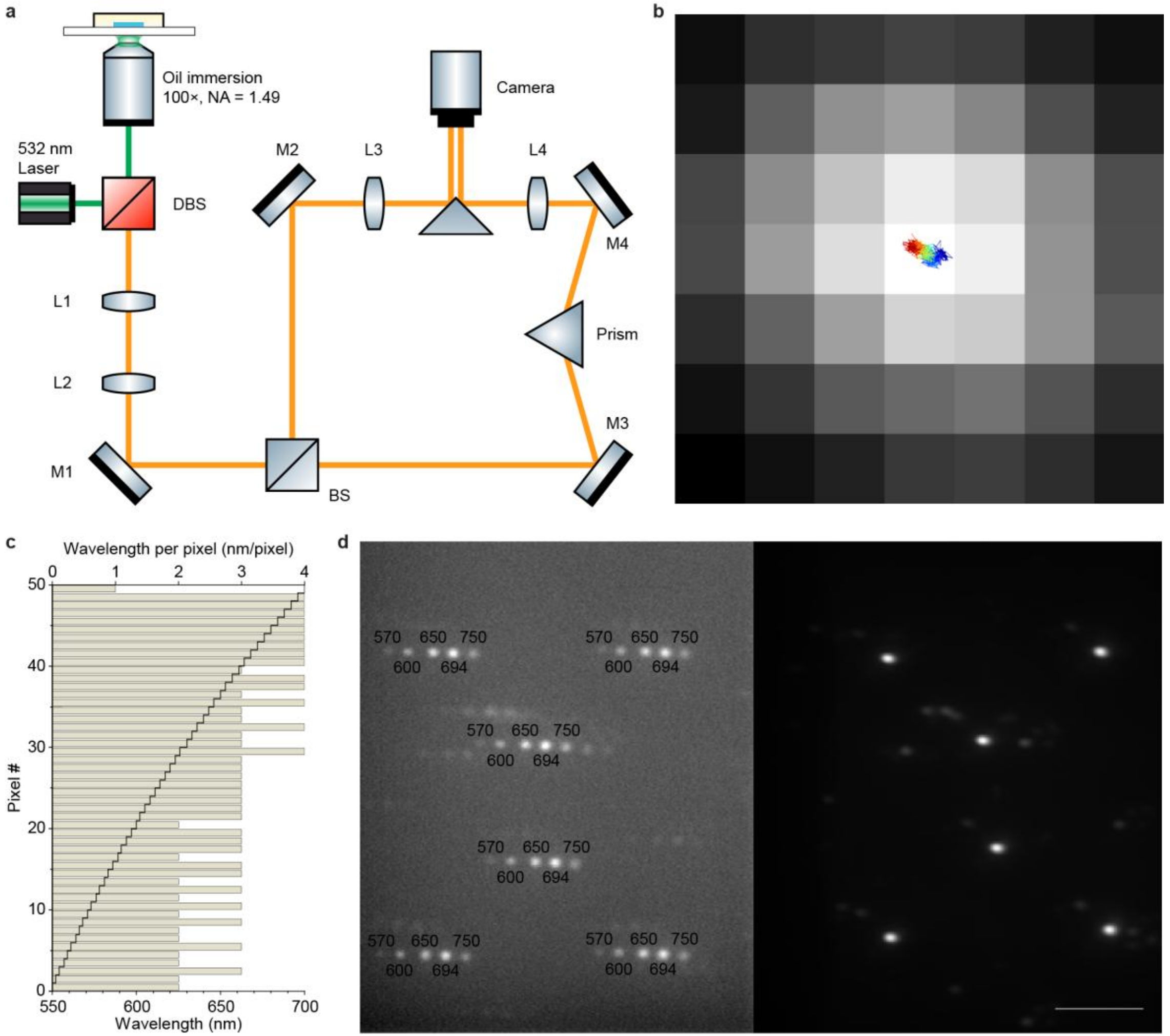

**Extended Data Fig. 3: Imaging setup and calibration.**

**a,** Schematic illustration of the imaging system. **b,** The movement trajectory of a calibration nanodiamond particle over 1000 frames, recorded at a frame rate of 100 ms. The trajectory points are colored according to frame number, transitioning from blue (initial frames) to red (final frames). The total drift distance is less than one pixel (88.88 nm), which is smaller than the deviation of the point spread function (PSF) of the emitters. **c,** Simulated wavelength calibration for an incident angle of 48.2° to the $CaF_2$ prism, using a wavelength resolution of 1 nm. At shorter wavelengths, each pixel corresponds to approximately 2–3 nm, while at longer wavelengths, each pixel represents about 4 nm. **d,** Composite image showing calibration particles (right) and their corresponding wavelength localizations (left) by stacking the calibration images together. A 2D regression model was used to correlate pixel positions with wavelength values across different spatial locations. Distinct color bars are used for the left and right part of the image to clearly differentiate spots from the wavelength domain (scale bar: 10 μm).

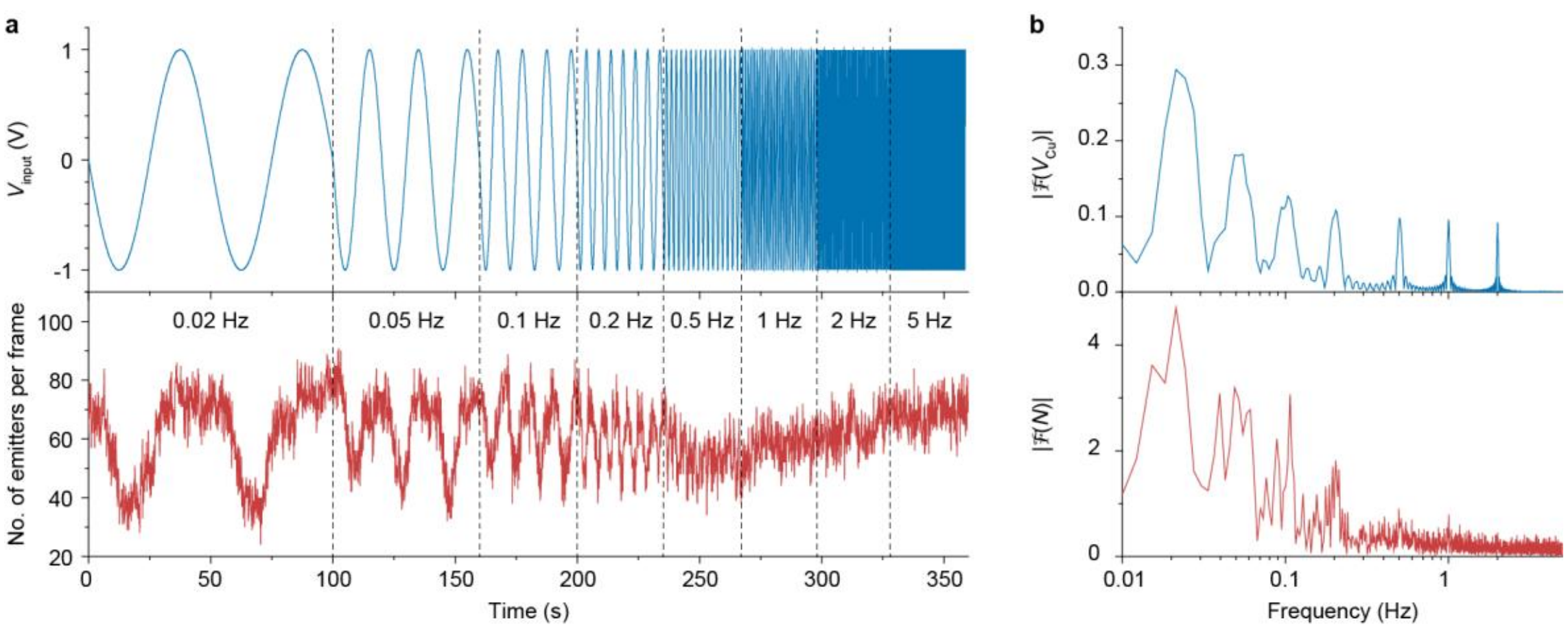

**Extended Data Fig. 4: Frequency domain analysis of ion dynamics under electrical field.** **a**, Voltage profile of the applied sine wave at different frequencies and the respective variation in number of emitters per frame. **b**, Fourier transform amplitudes for the applied voltage (input signal) and the number of emitters (output response) across varying frequencies.

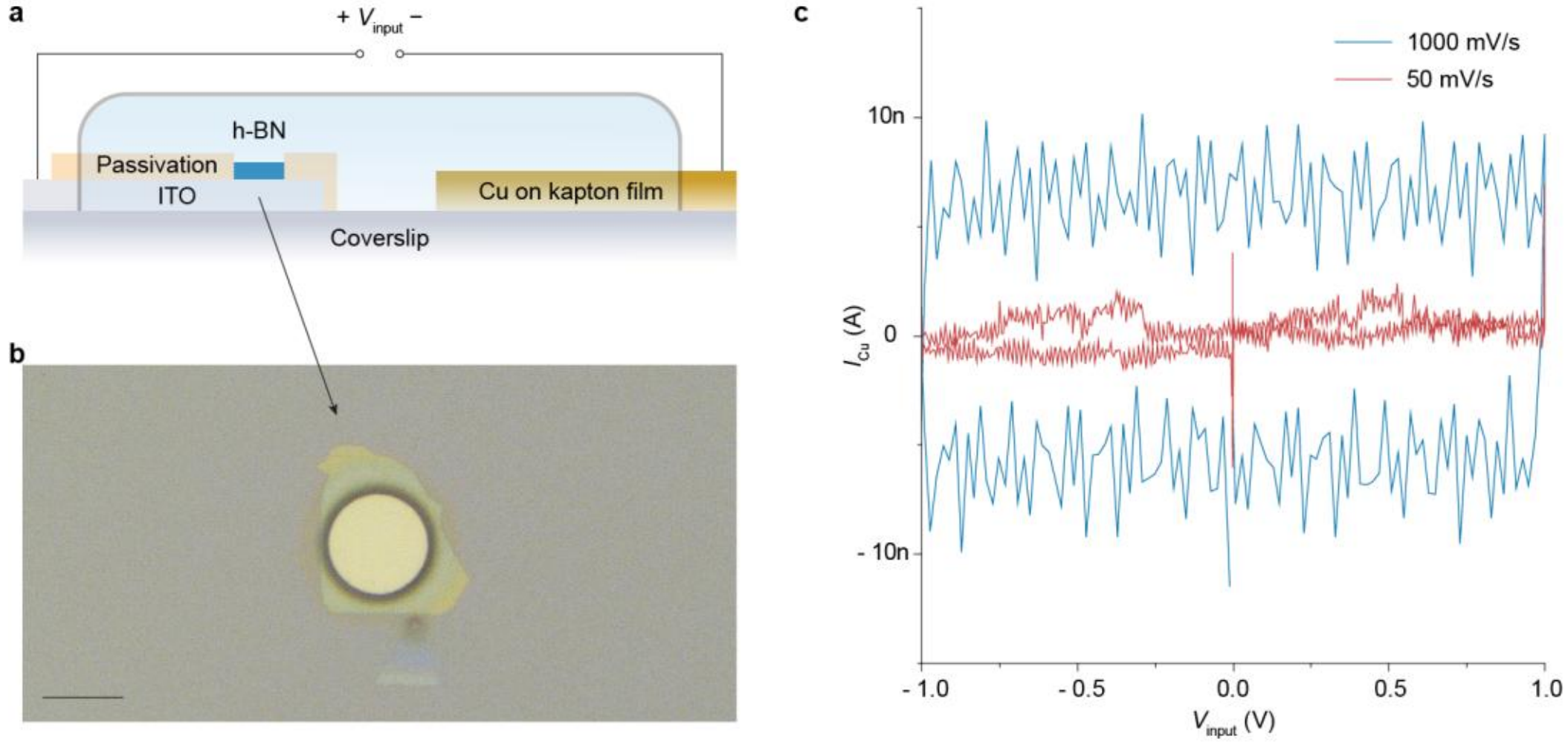

**Extended Data Fig. 5: Cyclic voltammetry scan of a passivated sample.**
**a-b**, Schematic (**a**) and optical photograph (**b**) under white light of a passivated device, where the substrate is coated with photoresist and a window is opened on the h-BN flake to expose it to the electrolyte (scale bar: 10 μm). **c**, Cyclic voltammetry (CV) scan of the device, the current at slow scan rates was below the noise level, while at fast scan rates the device exhibited capacitive behavior without any observable Faradaic current. The flake thickness is ~ 36 nm.

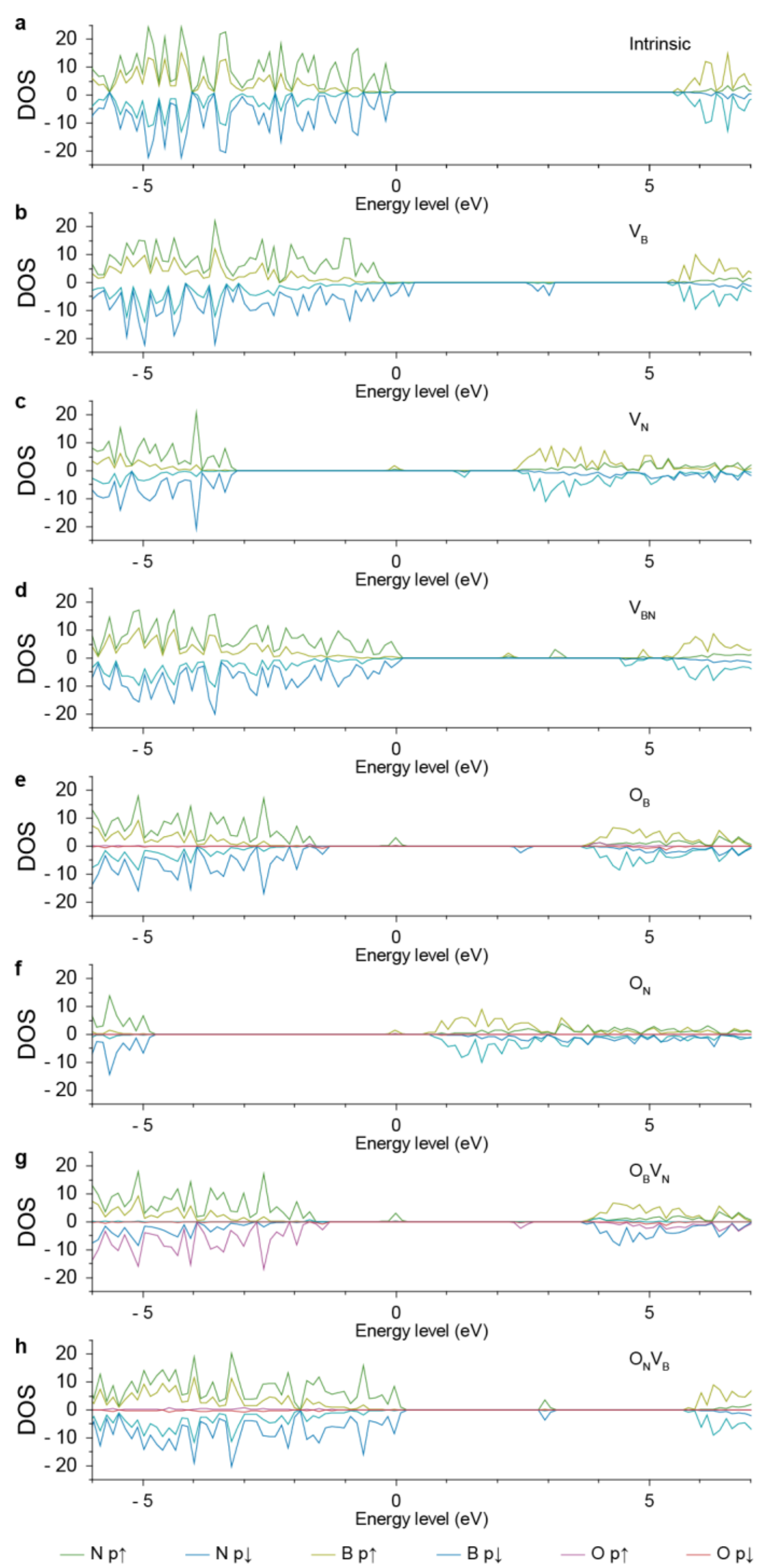

**Extended Data Fig. 6: Calculated density of states (DOS) using HSE06 hybrid functional.** **a,** intrinsic h-BN. The calculation is in good agreement with the measured experimental value of the indirect bandgap of 5.955 eV.[55] **b,** boron vacnacy ($V_B$). **c,** nitrogen vacancy ($V_N$). **d,** born and nitrogen pair vacancy ($V_{BN}$). **e,** Boron is substituted by an oxygen ($O_B$). **f,** Nitogen is substituted by an oxygen ($O_N$). **g,** Boron is substituted by an oxygen with a nitrogen vacancy ($O_BV_N$). **h,** Nitrogen is substituted by an oxygen with a boron vacancy ($O_NV_B$). Note that in this work, the excitation wavelength of 532 nm (2.33 eV) is the maximum excitation energy of electrons from the valance band to a defect state.

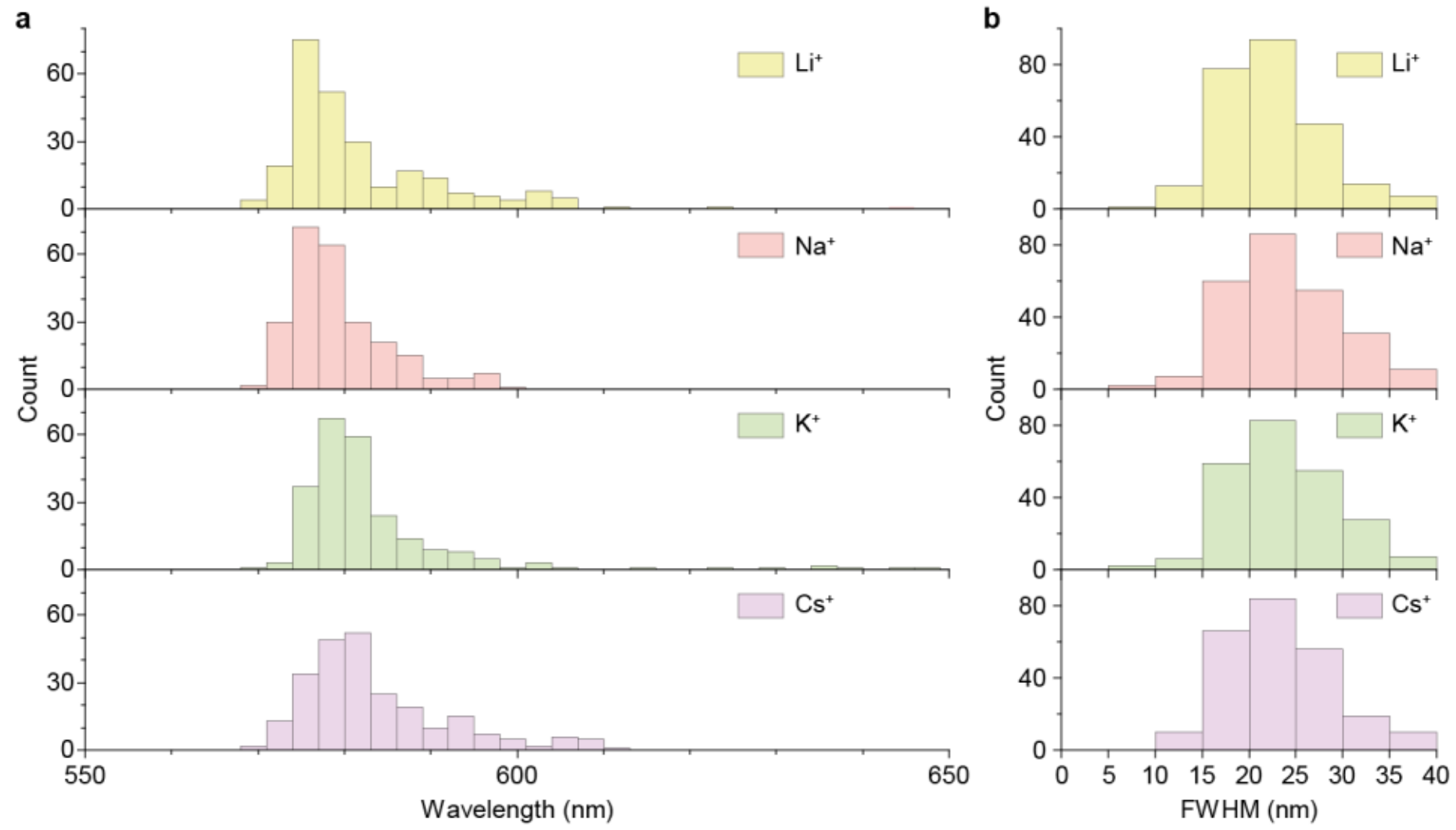

**Extended Data Fig. 7: Statistical results of h-BN single defect emissions with $Li^+$, $Na^+$, $K^+$, and $Cs^+$.**

**a,** Distribution of emission spectral peak centers from an identical flake in different aqueous solutions containing $Li^+$, $Na^+$, $K^+$, and $Cs^+$. The bins start at 550 nm with a width of 3 nm. **b,** Corresponding distributions of the full width at half maximum (FWHM) values for the same spectra in panel a**.** **c**, The fitted FWHM of the spectra in Fig. 4 **a-c**.

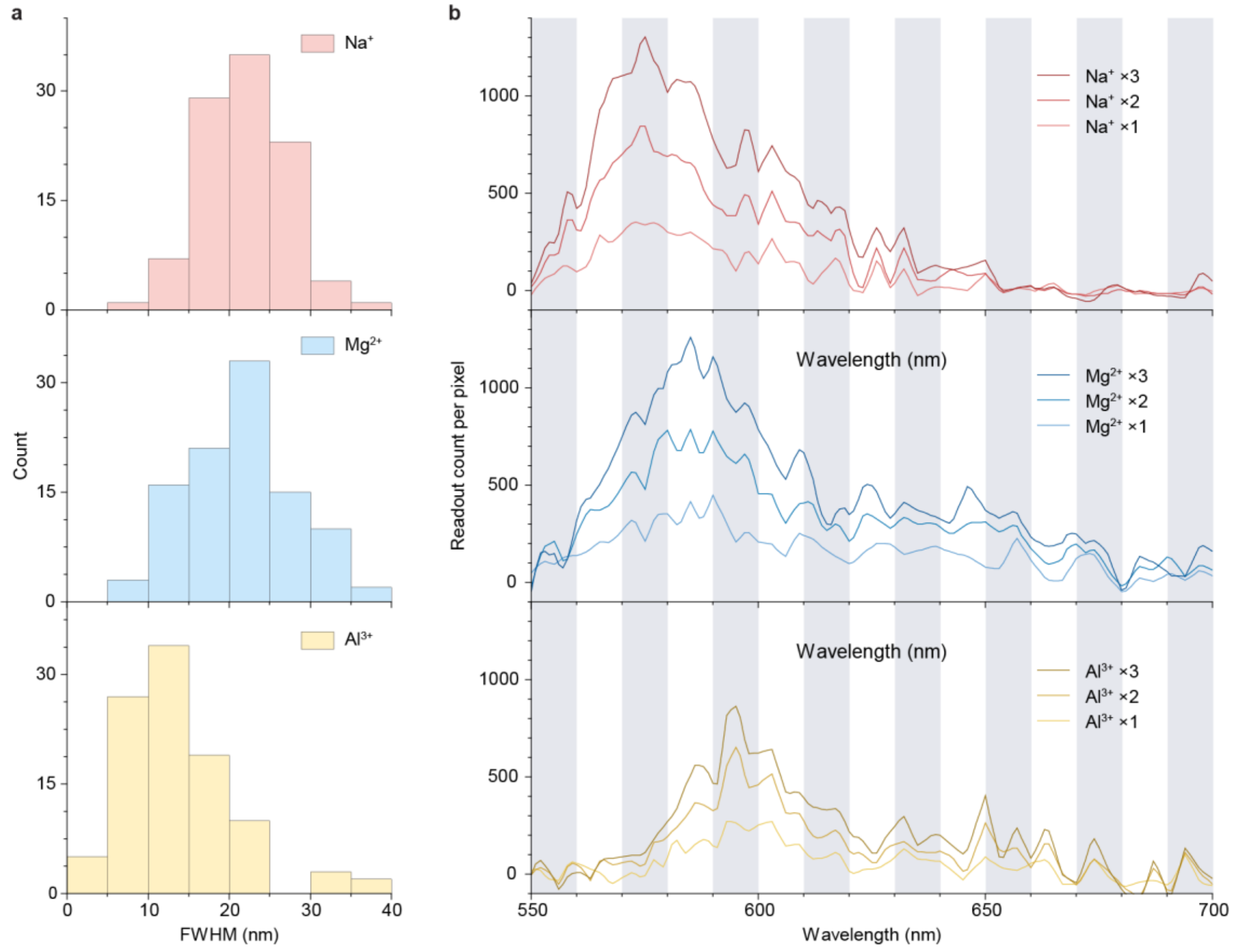

**Extended Data Fig. 8: Spectra analysis of h-BN single defect emissions with $Na^+$, $Mg^{2+}$, and $Al^{3+}$.**

**a**, The fitted FWHM of the spectra in Fig. 4 **a-c**. **b,** Accumulated spectra obtained by summing individual extracted spectra within each group in Fig. 4d–f. Increasing accumulation improves the signal-to-noise ratio (SNR).

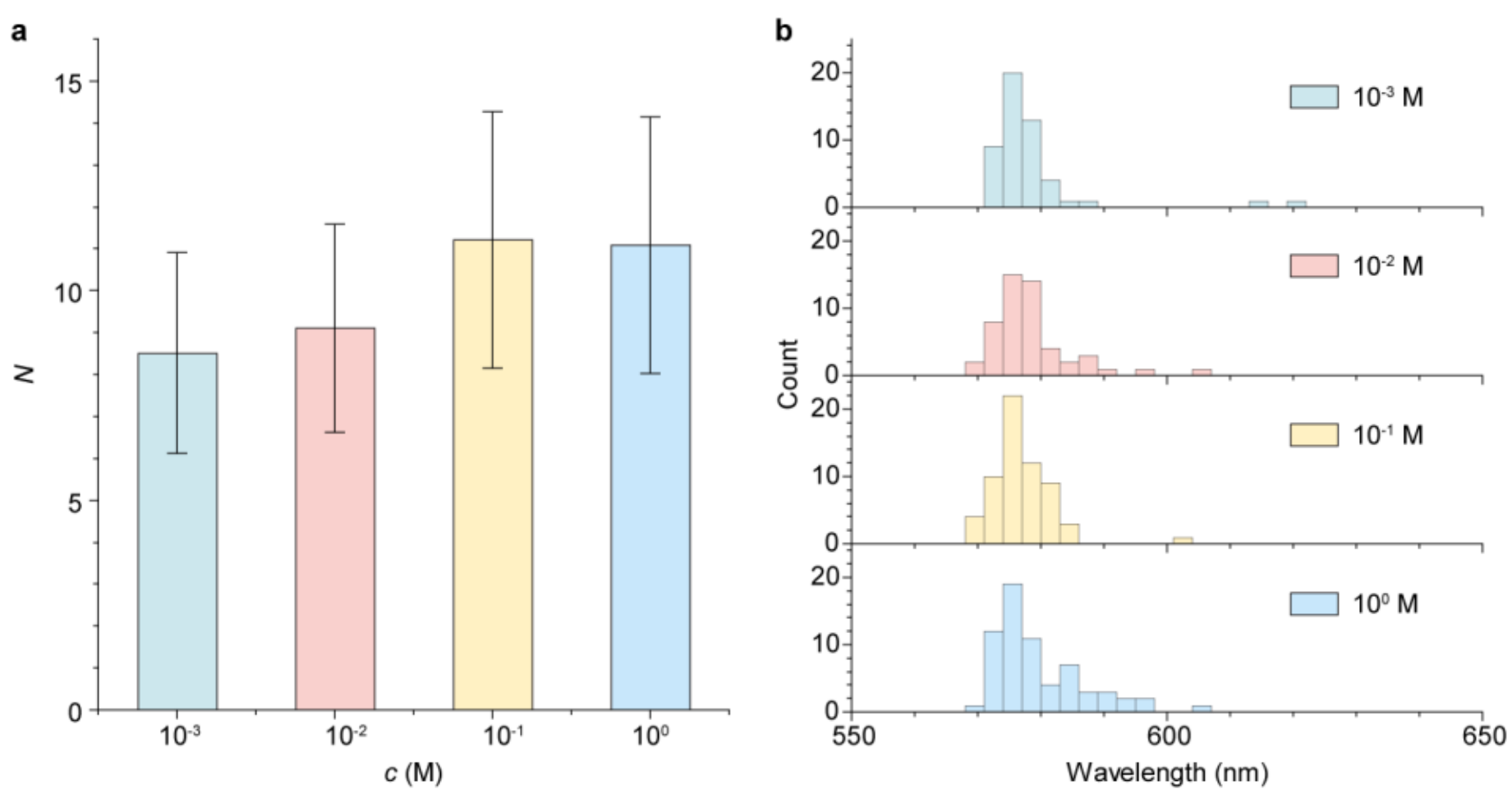

**Extended Data Fig. 9: Dependence of collected emission signal on concentration.**
**a,** Number of emitters per frame (20 ms exposure time) for the same h-BN flake imaged in aqueous solutions with varying $Li^+$ concentrations. The error bars represent the standard deviation. **b,** Distributions of emitter peak positions within the flake across different $Li^+$ concentrations. Bins are defined from 550 nm in 3 nm increments.

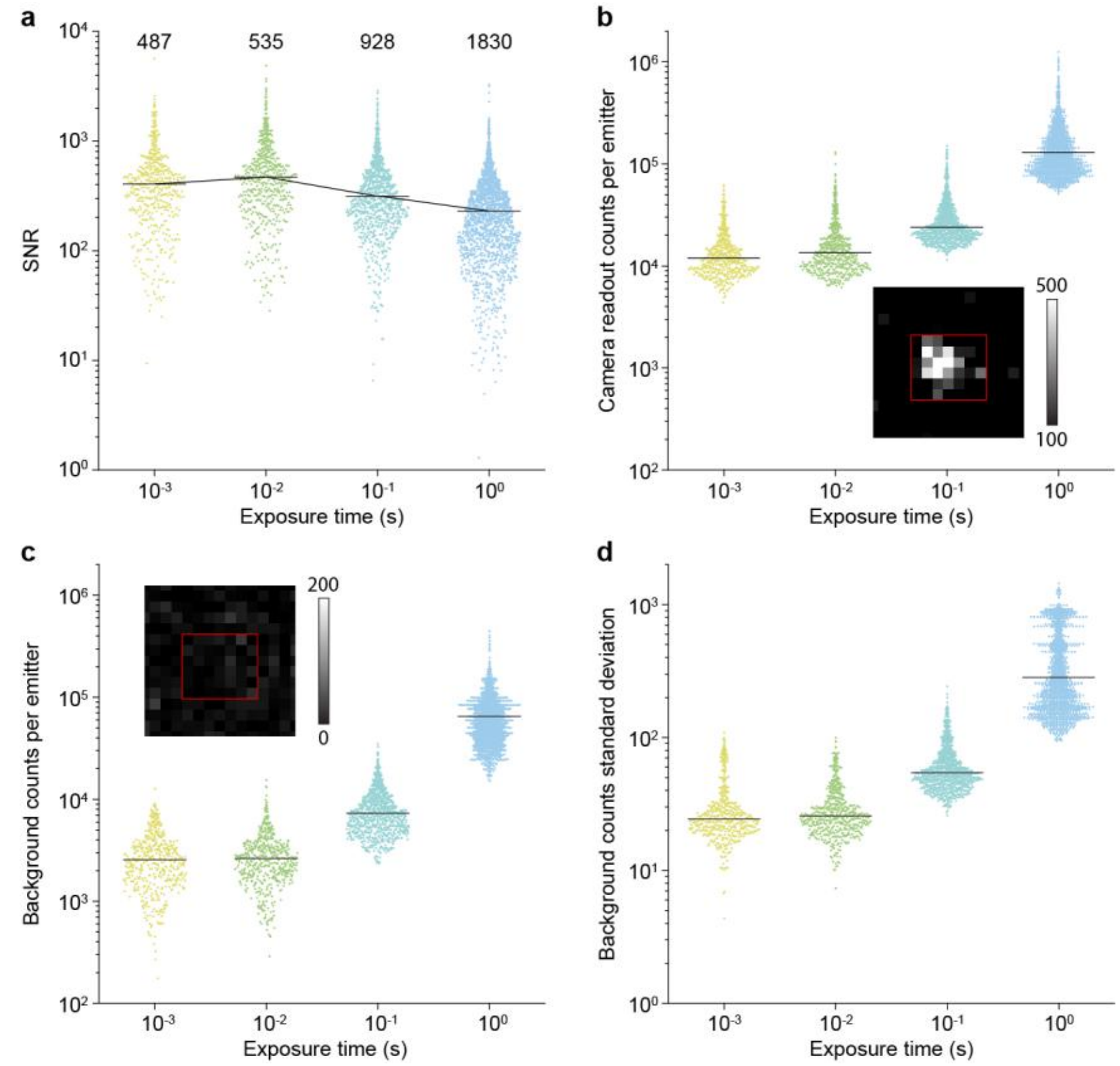

**Extended Data Fig. 10: Dependence of collected emission signal on exposure time.**
**a**, Distribution of the signal-to-noise ratio (SNR) for varying concentrations of $K^+$ ions. Details of the SNR calculations are provided in the Methods section. The numbers above the figure indicate the total number of emitters used for SNR analysis in 500 frames. The collision rate can be determined based on the exposure time of 10 ms. **b-d**, Distribution of camera readout counts (with EM gain set to 200 during data collection) for emitter (**b**), background (**c**), and standard deviation of the background photon counts (**d**) of each emitters. Examples of an emitter and its corresponding background are shown in the insets of panels **b** and **c**. The black lines are the median value. Here, only bright emitters, which account for the top 1% of pixel counts in the entire image, are included in the analysis. Dim emitters that cannot be used for spectral extraction are excluded from the plot. Projected pixel size is 88.9 nm. The flake thickness is ~ 19 nm.